\documentclass[aps,prd,notitlepage,amssymb,amsmath,floatfix,nofootinbib,superscriptaddress]{revtex4-1}

\usepackage{epsfig}
\usepackage{bm}
\usepackage{amssymb}
\usepackage{amsmath}
\usepackage{color}
\usepackage{slashed}
\usepackage{subfigure}
\usepackage{mathrsfs} 
\usepackage[colorlinks,linkcolor=blue,anchorcolor=black,citecolor=blue]{hyperref}
\allowdisplaybreaks[4]

\begin{document}

\title{The decay contribution to the parity-odd fragmentation functions}

\author{Yan-Lei Pan}
\affiliation{Institute of Frontier and Interdisciplinary Science, Key Laboratory of Particle Physics and Particle Irradiation (MOE), Shandong University, Qingdao, Shandong 266237, China}

\author{Kai-Bao Chen}
\email{chenkaibao19@sdjzu.edu.cn}
\affiliation{School of Science, Shandong Jianzhu University, Jinan, Shandong 250101, China}

\author{Yu-Kun Song}
\email{sps\_songyk@ujn.edu.cn}
\affiliation{School of Physics and Technology, University of Jinan, Jinan, Shandong 250022, China}

\author{Shu-Yi Wei}
\email{shuyi@sdu.edu.cn}
\affiliation{Institute of Frontier and Interdisciplinary Science, Key Laboratory of Particle Physics and Particle Irradiation (MOE), Shandong University, Qingdao, Shandong 266237, China}

\begin{abstract}
Parity violation in QCD is a consequence of the so-called QCD $\theta$-vacuum. As a result, parity-odd fragmentation functions are introduced and they bring in new observables in the back-to-back dihadron productions in $e^+e^-$-annihilation experiments~\cite{Kang:2010qx}. The experimental measurements on the corresponding parity-odd fragmentation functions can shed light on the local CP violation effect in QCD. On the other hand, the weak interaction also violates the parity symmetry. Therefore, the weak decay of heavier hadrons can also contribute to the parity-odd effects in fragmentation functions. In this paper, we investigate the weak decay contribution to these parity-odd fragmentation functions and compute their contribution to these new observables. In principle, the decay contribution should/can be excluded in the theoretical analysis and experimental measurements. However, this is usually not the common practice so far. Furthermore, in light of that the value of the $\theta$-parameter is extremely small ($\theta < 3 \times 10^{-10}$), the weak-parity-violating contributions become an important background in identifying the strong-parity-violating term. In this paper, we focus on the weak decay contribution of parity-odd fragmentation functions and demonstrate their sizable contribution in a numerical estimate.
\end{abstract}

\maketitle

\section{Introduction}

The modern concept of fragmentation function is established on the foundation of the QCD factorization theorem. Depending on the spin of the final state hadron and the factorization scheme, various fragmentation functions are introduced to fully describe the hadronization process of a high-energy parton \cite{Metz:2016swz, Chen:2023kqw}, which in return induce numerous intriguing phenomena in different high-energy collisions. 

Due to the non-perturbative nature, it is still not feasible to perform a first principle calculation of these quantities. Even the mighty lattice QCD approach cannot propose a workable-in-principle solution yet. Therefore, most theoretical studies focus on investigating the qualitative features. According to the power counting, fragmentation functions can be cast into leading twist and higher twist categories. While the leading twist ones enjoy clear physical interpretations, the higher-twist fragmentation functions are interference terms that cannot be perceived as number densities. Nevertheless, after convoluting with the corresponding higher twist hard part, they become parts of the differential cross section and therefore lead to measurable effects. In the past decades, a lot of theoretical efforts have been carried out towards the leading and higher twist components of spin-half and spin-one hadrons~\cite{Ji:1993vw, Mulders:1995dh, Boer:1997mf, Boer:1997qn, Boer:1999uu, Bacchetta:2000jk, Pitonyak:2013dsu, Wei:2013csa, Wei:2014pma, Chen:2015ora, Chen:2016moq, Chen:2016iey, Yang:2016qsf, Yang:2017sxz, Gamberg:2018fwy, Anselmino:2019cqd, Chen:2020pty, Chen:2021hdn, Chen:2021zrr, Jiao:2022gzu, Zhao:2022lbw, Song:2023ooi}, which established the connection between fragmentation functions and experimental observables. 

On the quantitative study side, fragmentation functions can only be either extracted from experimental data (e.g.,~\cite{Binnewies:1994ju, deFlorian:1997zj, Bourhis:2000gs, deFlorian:2007aj, deFlorian:2007ekg, Anselmino:2007fs, Hirai:2007cx, Albino:2008fy, Anselmino:2013vqa, Anselmino:2013lza, deFlorian:2014xna, Anselmino:2015sxa, Anselmino:2015fty, deFlorian:2017lwf}) or computed in the context of models~\cite{Nzar:1995wb, Jakob:1997wg, Bacchetta:2001di, Metz:2002iz, Bacchetta:2003xn, Gamberg:2003eg, Amrath:2005gv, Lu:2015wja, Yang:2016mxl, Yang:2017cwi, Wang:2018wqo, Xie:2022lra}. A comprehensive review of the experimental measurements and model calculations is given in Sections 4 and 6 of Ref.~\cite{Metz:2016swz}. While the quantitative extraction of the unpolarized fragmentation functions is already quite mature~\cite{Binnewies:1994ju, Bourhis:2000gs, Kretzer:2000yf, Hirai:2007cx, Albino:2008fy, Aidala:2010bn, dEnterria:2013sgr, deFlorian:2014xna, deFlorian:2017lwf}, that of the spin-dependent ones is still preliminary due to the limited resource. Only several studies on the spin-dependent fragmentation functions are currently available~\cite{deFlorian:1997zj, Anselmino:2000vs, Anselmino:2019cqd, DAlesio:2020wjq, Callos:2020qtu, Gamberg:2021iat, Chen:2021hdn, DAlesio:2022brl, DAlesio:2023ozw}. This situation will be significantly improved by the renewed attention from the Belle experiment~\cite{Belle-II:2018jsg,Belle:2018ttu} and future Electron-Ion Collider~\cite{Boer:2011fh, Accardi:2012qut, AbdulKhalek:2021gbh}. Particularly, recent results from the Belle experiment~\cite{Belle:2018ttu} on the spontaneous transverse polarization of the $\Lambda$ hyperon have sparked several theoretical studies from different angles~\cite{Gamberg:2018fwy, Anselmino:2019cqd, DAlesio:2020wjq, Callos:2020qtu, Li:2020oto, Gamberg:2021iat, Chen:2021hdn, Chen:2021zrr, DAlesio:2022brl, DAlesio:2023ozw}. 

Fragmentation functions describe the hadron production from a fragmenting quark. This process is dominated by the strong interaction which is usually considered to preserve the charge conjugation and parity symmetries. Therefore, parity-violating (P-odd) effects are prohibited. However, a pioneer work~\cite{Kang:2010qx} argued that QCD can break the local CP-symmetry through the $\theta$-vacuum. As a consequence, P-odd fragmentation functions can be introduced. The experimental measurements of these functions can be utilized as a probe to the $\theta$ parameter. Further studies of P-odd fragmentation functions are also presented in Refs.~\cite{Yang:2019rrn, Yang:2019gdr}. Moreover, in the context of parton distribution functions, the P-odd effect has also been extensively investigated in Refs.~\cite{Yang:2019hxu, Bacchetta:2023hlw}.

In the decomposition of fragmentation functions, parity conservation is usually employed to constrain possible structures. We use the leading twist decomposition as an example to illustrate this point. In the collinear factorization, we can only define three prompt fragmentation functions for the spin-$1/2$ hadron production, namely the unpolarized fragmentation function, the longitudinal spin transfer, and the transverse spin transfer. The unpolarized fragmentation function, usually denoted as $D_{1}$, represents the number density of producing unpolarized hadrons from unpolarized quarks. The longitudinal/transverse spin transfer ($G_{1L}$/$H_{1T}$) is the number density of producing longitudinally/transversely polarized hadrons from longitudinally/transversely polarized quarks. The other combinations like $D_{1L}^{\rm PV}$ which represents the number density of producing longitudinally polarized hadrons from unpolarized quarks and $G_{1}^{\rm PV}$ which represents that of producing unpolarized hadrons from longitudinally polarized quarks are exact zero since they break parity symmetry. Once the QCD $\theta$-vacuum is taken into account, the onset of parity violation leads to non-zero contributions to the P-odd fragmentation functions. The experimental measurements of P-odd fragmentation functions can be utilized to constrain the tiny $\theta$-parameter. 

Besides the QCD $\theta$-term (denoted bellow as the strong-parity-violating term), the weak interaction also breaks the parity conservation. Although the hadronization process is dominated by the strong interaction, the weak interaction can also contribute through the decay channels. The decay contribution is usually not excluded in the experimental measurements (see e.g.,~\cite{ALEPH:1996oew, OPAL:1997oem}). Therefore, in addition to the prompt hadron production in the hadronization process, the decay of heavier hadrons can also contribute to the total production of the final state hadrons. Besides modifying the production rate of hadrons, weak decays also contribute to the P-odd fragmentation functions. To acquire a precise constraint on the tiny $\theta$-parameter by measuring P-odd fragmentation functions, we must understand the weak decay contribution part first. This is the main goal of our study.

In this paper, we will not study the contribution of P-odd fragmentation functions arising from the strong-parity-violating term. Instead, we only focus on the contribution from weak decays. We present our numerical estimates of their contribution to the P-odd fragmentation functions and predict corresponding observables. We demonstrate that the weak decay contribution is sizable. Therefore, a rigorous treatment of the weak decay contribution is indeed essential to correctly identify the strong-parity-violating effect and constrain the tiny QCD $\theta$-parameter. 

This paper is organized as follows. In Sec. II, we present the general formula to study the decay contribution of fragmentation functions. In Sec. III, we present our numerical estimates of P-odd fragmentation functions. In Sec. IV, we make predictions for the $\Lambda$ longitudinal polarization and the suppression of di-pion production in $e^+e^-$ annihilations. A summary is given in Sec. V.

\section{General formula for decay contributions}

Different hadrons decay through different channels. In this work, we propose a simplification that relates a multi-body decay process to a two-body decay process. For a multi-body decay process $H \to h +X$ where $H$ is the parent hadron, $h$ is the daughter hadron of interest, and $X$ is a collection of other particles, our simplification is to regard $X$ as a pseudo-particle with mass being denoted as $m_X$. Unlike the two-body decay process where $X$ is a real particle, $m_X$ in the multi-body decay process is not a fixed number. For any multi-particle decay process, we can numerically obtain the distribution of $m_X$ which is denoted by $f(m_X)$. The contribution from multi-particle decay is given by an integral of $m_X$ in the allowed kinematic region. In light of this, a two-body decay process can be viewed as a special case of a multi-body decay process where $f(m_X)= \delta(m_X-m_0)$ with $m_0$ the mass of the real particle. Finally, we arrive at
\begin{align}
\frac{dN^{\text{mb}}}{dPS} = \int dm_X f(m_X) \frac{dN^{\text{2b}}}{dPS}, \label{eq:manybody}
\end{align}
where ${dN^{\text{mb}}}/{dPS}$ is the differential decay distribution of a general many-body decay and ${dN^{\text{2b}}}/{dPS}$ is that of a two-body decay process with $X$ being regarded as a pseudo-particle. Apparently ${dN^{\text{2b}}}/{dPS}$ depends on $m_X$. 

In the following subsections, we consider three special cases that will be used in the numerical evaluation and lay out the full formula for a two-body process. The extension to a multi-body decay process is straightforward equipped with Eq.~(\ref{eq:manybody}).

\subsection{Spin-$1/2$ hadron decays into spin-$1/2$ hadron}

To be more specific, we first consider the $H \to h + X$ decay process with both $H$ and $h$ being spin-$1/2$ hadrons. We use ${\cal D}^{H}_{q} (\lambda_q, \lambda_H; z', p_T')$ to denote the helicity-dependent fragmentation function of the parent hadron $H$ and ${\cal D}^{h,H}_q (\lambda_q, \lambda_h; z, p_T)$ to denote the decay contribution to that of the daughter hadron $h$ with $\lambda_q$ and $\lambda_{h/H}$ being the helicities of the quark and the corresponding hadron. $X$ is a collection of all the other unmeasured particles. Here, $z$ and $z'$ are the longitudinal momentum fractions of daughter and parent hadrons respectively, while $p_T$ and $p_T'$ are the corresponding transverse momenta with respect to the quark momentum. 

For the parent hadron $H$ production, we neglect decay contributions and only consider the prompt production contribution, which is parity invariant. Therefore, we only have the following two possible structures as long as we only consider the longitudinal polarizations of the quark and the hadron,
\begin{align}
{\cal D}^{H}_{q} (\lambda_q, \lambda_H; z, p_T)
= {\cal D}^{H,{\rm dir}}_{q} (\lambda_q, \lambda_H; z, p_T)
= D_{1,q}^{H,{\rm dir}} (z, p_T) + \lambda_q \lambda_H G_{1L,q}^{H,{\rm dir}} (z, p_T),
\end{align}
where $D_{1,q}^{H,{\rm dir}} (z, p_T)$ and $G_{1L,q}^{H,{\rm dir}} (z, p_T)$ are the unpolarized fragmentation function and the longitudinal spin transfer for the prompt production respectively. 

Once the transverse polarizations of the quark and/or the hadron are taken into account, the other six leading twist fragmentation functions can also be defined this way. These structures are beyond the scope of our current study. In this paper, we only study the longitudinal polarizations of both the quark and the hadron. 

Since we are going to consider the general decay kinematics, we keep the transverse momentum dependence at the current stage. These transverse-momentum-dependent fragmentation functions are related to their collinear partners by
\begin{align}
& D_{1,q}^{H,{\rm dir}} (z) = \int d^2 p_T D_{1,q}^{H,{\rm dir}} (z, p_T),
& G_{1L,q}^{H,{\rm dir}} (z) = \int d^2 p_T G_{1L,q}^{H,{\rm dir}} (z, p_T).
\end{align}

Thanks to the weak decay contribution, the total fragmentation function is allowed to break the parity symmetry. In general, we have
\begin{align}
{\cal D}^{h}_q (\lambda_q, \lambda_h; z, p_T) = 
D_{1,q}^{h} (z, p_T) + \lambda_q \lambda_h G_{1L,q}^{h} (z, p_T) + \lambda_h D_{1L,q}^{{\rm PV},h} (z,p_T) + \lambda_q G_{1,q}^{{\rm PV},h} (z,p_T).
\end{align}
Here, $D_{1L,q}^{{\rm PV},h}$ and $G_{1,q}^{{\rm PV},h}$ represent the number densities of producing longitudinally polarized hadrons from unpolarized quarks and of producing unpolarized hadrons from longitudinally polarized quarks respectively. They are apparently P-odd. Notice that we have employed a different naming system with that in Ref.~\cite{Kang:2010qx}, where $\tilde D$ in their paper is denoted as $G_{1,q}^{{\rm PV},h}$ in this paper. In this work, we adopt the following convention: (i) $D$ denotes a fragmentation function of an unpolarized quark and $G$ denotes that of a longitudinally polarized quark. (ii) The subscript $L$ is used to indicate the fragmentation function for the production of longitudinally polarized hadrons. (iii) The ``PV'' in the superscript of $D$ or $G$ is employed to represent the parity violation. We do not consider the $\theta$-vacuum term in the QCD Lagrangian so that the P-odd fragmentation functions solely arise from the weak decay. 

Convoluting the parent hadron fragmentation function with the decay kernel function, we arrive at
\begin{align}
{\cal D}^{h,H}_q (\lambda_q, \lambda_h; z, p_T)
= \sum_{\lambda_H} \int dz' d^2p_T' \frac{dN (\lambda_h, \lambda_H)}{dz d^2p_T} {\cal D}^{H}_{q} (\lambda_q, \lambda_H; z', p_T'),
\end{align}
where ${\cal D}^{h,H}$ denotes the decay contribution of the parent hadron $H$ to the fragmentation function of $h$ and ${dN (\lambda_h, \lambda_H)}/{dz d^2p_T}$ is the number density for a parent hadron $H$ with helicity $\lambda_H$ and four-momentum being specified by $z'$ and $p_T'$ to produce a daughter hadron $h$ with helicity $\lambda_h$ and four-momentum being labeled by $z$ and $p_T$. Notice that the helicity direction is frame-dependent for massive particles. Throughout this paper, the helicity is always defined in the lab frame. The decay contributions to spin-dependent fragmentation functions thus depend on the parton energy. Unlike the factorization scale dependence which obeys the QCD evolution equation, this dependence on the parton energy arises from the kinematics rather than from the factorization theorem. 

For a two-body decay process, the differential distribution $dN/dzd^2p_T$ is given by
\begin{align}
\frac{dN(\lambda_h,\lambda_H)}{dzd^2p_T} 
= & \frac{1}{8\pi} \frac{2M_H}{z|\bm{p}_h^*|} 
\delta [(p_H-p_h)^2-M_X^2]
\Biggl[ 
1 + \gamma \lambda_H\lambda_h \bm{\omega}_i \cdot \bm{\omega}_f + (1-\gamma) \lambda_H \lambda_h 
\left(\bm{\omega}_i \cdot \frac{\bm{p}_h^*}{|\bm{p}_h^*|} \right)
\left(\bm{\omega}_f \cdot \frac{\bm{p}_h^*}{|\bm{p}_h^*|} \right)
\nonumber \\
&  
+ \alpha (\lambda_H \bm{\omega}_i \cdot \frac{\bm{p}_h^*}{|\bm{p}_h^*|} + \lambda_h \bm{\omega}_f \cdot \frac{\bm{p}_h^*}{|\bm{p}_h^*|}) + \beta \lambda_H \lambda_h \frac{\bm{p}_h^*}{|\bm{p}_h^*|} \cdot (\bm{\omega}_f \times \bm{\omega}_i)
\Biggr], \label{eq:onehalftoonehalf}
\end{align}
where $\alpha$, $\beta$ and $\gamma$ are constant parameters, $M_H$ is the mass of $H$, $\bm{p}_h^*$ is the three-momentum of $h$ in the $H$-rest frame, $p_H$ and $p_h$ are the four-momenta of $H$ and $h$ in the lab frame, $\bm{\omega}_i = \bm{p}_H/|\bm{p}_H|$ is the helicity direction of $H$ and $\bm{\omega}_f$ is the helicity direction of $h$ in the $H$-rest frame. Due to the Wick rotation, the helicity direction $\boldsymbol{\omega}_f$ is not along $\bm{p}_h^*$. It is given by
\begin{align}
\bm{\omega}_f = \frac{1}{M_H}\left[ \frac{m_h E_H + M_H E_h}{(E_h^* + m_h)|\bm{p}_h|} \bm{p}_h^* + \frac{m_h}{|\bm{p}_h|} \bm{p}_H \right],
\end{align}
where, $E_{H/h}$ is the energy of $H/h$ in the lab frame, $\bm{p}_{H/h}$ is the three momentum of $H/h$ in the lab frame, $E_h^*$ is the energy of $h$ in the $H$-rest frame and $m_h$ is the mass of $h$. Despite the involved expression, the angle between $\bm{\omega}_f$ and $\bm{p}_h^*/|\bm{p}_h^*|$ is just the well-known Wick-angle $\theta_{\rm Wick}$. Taking the $m_h \to 0$ limit, we find that $\cos\theta_{\rm Wick} = \bm{\omega}_f \cdot \bm{p}_h^* / |\bm{p}_h^*| =1$. This means that the helicity of a massless particle does not change under Lorentz boosts.

We only consider the prompt production for the parent hadron $H$. Therefore, only parity-even terms survive. Benefiting from the number density interpretation of leading twist fragmentation functions, it is straightforward to obtain the decay contribution to the $h$ fragmentation function as
\begin{align}
&
D_{1,q}^{h,H} (z, p_T) = \int dz' d^2 p_T' D_{1,q}^{H} (z',p_T') \times \frac{1}{2} \sum_{\lambda_h,\lambda_H} \frac{dN (\lambda_h,\lambda_H)}{dzd^2p_T}
,\\
&
G_{1L,q}^{h,H}(z,p_T) = 
\int dz' d^2 p_T' G_{1L,q}^{H} (z', p_T') \times \frac{1}{2}
\sum_{\lambda_H} \lambda_H \left[ \frac{dN (+,\lambda_H)}{dzd^2p_T} - \frac{dN (-,\lambda_H)}{dzd^2p_T} \right]
,\\
&
D_{1L,q}^{{\rm PV},h,H} (z,p_T) = 
\int dz' d^2 p_T' D_{1,q}^{H} (z', p_T') \times \frac{1}{2}
\sum_{\lambda_H}\left[ \frac{dN (+,\lambda_H)}{dzd^2p_T} - \frac{dN (-,\lambda_H)}{dzd^2p_T} \right]
,\\
&
G_{1,q}^{{\rm PV},h,H} (z,p_T) = 
\int dz' d^2 p_T' G_{1L,q}^{H} (z', p_T') \times \frac{1}{2}
\sum_{\lambda_h} \left[ \frac{dN (\lambda_h,+)}{dzd^2p_T} - \frac{dN (\lambda_h,-)}{dzd^2p_T} \right]
.
\end{align}
Furthermore, we can integrate over $d^2 p_T$ and obtain
\begin{align}
&
D_{1,q}^{h,H} (z) = \frac{M_H}{2|\bm{p}_h^*|} \int \frac{dz'}{z'} d^2 p_T' D_{1,q}^H (z', p_T') K_{U\to U},
\\
&
G_{1L,q}^{h,H} (z) = \frac{M_H}{2|\bm{p}_h^*|} \int \frac{dz'}{z'} d^2 p_T' G_{1L,q}^H (z', p_T') K_{L\to L},
\\
&
D_{1L,q}^{{\rm PV},h,H} (z) = \frac{M_H}{2|\bm{p}_h^*|} \int \frac{dz'}{z'} d^2 p_T' D_{1,q}^H (z', p_T') K_{U\to L},
\label{eq:d1ltilde}
\\
&
G_{1,q}^{{\rm PV},h,H} (z) = \frac{M_H}{2|\bm{p}_h^*|} \int \frac{dz'}{z'} d^2 p_T' G_{1L,q}^H (z', p_T') K_{L\to U},
\end{align}
where the kernel functions are given by
\begin{align}
& K_{U\to U} = 1,
\\
& K_{L\to L} = \gamma \frac{E_H^2 m_h^2 + E_h^2 M_H^2 - E_H E_h M_H (E_h^*-m_h) - m_h M_H^2(E_h^* + m_h)}{M_H (E_h^*+m_h)|\bm{p}_H||\bm{p}_h|} 
\nonumber\\
& \phantom{XXXXx}
+ (1-\gamma) \frac{M_H E_h - E_H E_h^*}{|\bm{p}_H||\bm{p}_h^*|} \frac{M_H E_h E_h^* - E_H m_h^2}{M_H |\bm{p}_h||\bm{p}_h^*|},
\\
& K_{U\to L} = \alpha \frac{M_H E_h E_h^* - E_H m_h^2}{M_H |\bm{p}_h||\bm{p}_h^*|},
\\
& K_{L\to U} = \alpha \frac{M_H E_h - E_H E_h^*}{|\bm{p}_H||\bm{p}_h^*|}.
\end{align}
Although the above expressions are complicated, the physical interpretations of these kernel functions are straightforward. For instance, $K_{U \to L} = \alpha \cos \theta_{\rm Wick}$ with $\theta_{\rm Wick}$ being the Wick angle. $K_{L \to U} = \alpha \cos \theta^*$ with $\theta^*$ the polar angle of $h$-momentum in the $H$-rest frame. $K_{L \to L} = \cos \theta^* \cos \theta_{\rm Wick} + \gamma \sin \theta^* \sin \theta_{\rm Wick}$. Furthermore, we have also run a simple Monte-Carlo simulation to perform the cross-check. Numerical evaluation with the above kernel functions can exactly reproduce the simulation results.

\subsection{Spin-$3/2$ hadron decays into spin-$1/2$ hadron}

The polarization of a spin-$3/2$ hadron consists of a lot of modes. Each polarization mode is associated with multiple fragmentation functions~\cite{Zhao:2022lbw} describing the production of spin-$3/2$ hadrons. Unfortunately, all of the spin-dependent fragmentation functions are poorly studied numerically so far. In this study, we average over the polarization of the parent hadron and only study the induced longitudinally polarization of the final state $\Lambda$-hyperon, and we arrive at  
\begin{align}
\frac{dN(\lambda_h)}{dzd^2p_T} 
= & \frac{1}{8\pi} \frac{2M_H}{z|\bm{p}_h^*|} 
\delta [(p_H-p_h)^2-M_X^2]
\Biggl[ 
1 + \alpha \lambda_h \bm{\omega}_f \cdot \frac{\bm{p}_h^*}{|\bm{p}_h^*|} 
\Biggr],
\end{align}
where $\alpha$ is the decay parameter of spin-$3/2$ hadron measuring the magnitudes of the parity violation in the weak decay. The rest of the calculation coincides with that for the spin-$1/2$ to spin-$1/2$ decay.

\subsection{Spin-$1/2$ hadron decays into spin-$0$ hadron}

Besides the P-odd fragmentation functions of the $\Lambda$-hyperon, we also want to investigate those of pions. Most of the vector mesons decay through the strong interaction and therefore do not bring in any P-odd effect. Effectively, we only need to consider the weak decay of spin-$1/2$ hadrons. It is straightforward to obtain the decay kernel function for this case from Eq.~(\ref{eq:onehalftoonehalf}) by setting $\lambda_h=0$ and $\bm{\omega}_f = 0$. We obtain
\begin{align}
\frac{dN(\lambda_H)}{dzd^2p_T} 
= & \frac{1}{8\pi} \frac{2M_H}{z|\bm{p}_h^*|} 
\delta [(p_H-p_h)^2-M_X^2]
\Biggl[ 
1 + \alpha \lambda_H \bm{\omega}_i \cdot \frac{\bm{p}_h^*}{|\bm{p}_h^*|} 
\Biggr]. \label{eq:onehalftozero}
\end{align}
When the formula is served like Eq.~(\ref{eq:onehalftozero}), the sign of the $\alpha$ parameter depends on which hadron is the one of interest. For instance, in the $\Lambda \to p + \pi^-$ decay process, $\alpha_p = - \alpha_\pi$.

\section{Numerical results of P-odd fragmentation functions}

In the previous section, we have demonstrated the emergence of P-odd fragmentation functions resulting from weak decays. In this section, we present our numerical estimates. 

As shown in Eq.~(\ref{eq:d1ltilde}), $D_{1L}^{{\rm PV}}$ can be written as a convolution of the unpolarized parent hadron fragmentation function with the $K_{U\to L}$ kernel function. Therefore, to numerically estimate $D_{1L}^{\rm PV}$, we need to parameterize the unpolarized fragmentation functions of all hadrons that can decay into the daughter hadron. For the $\Lambda$ production, we only need to consider the weak decay of $\Xi$ and $\Omega^-$. While the other hadrons can also decay into $\Lambda$ through strong or electromagnetic interaction, they do not contribute to P-odd effects. They only contribute to the decay contribution of the unpolarized fragmentation function. The contribution is included in our numerical calculation, but the results are not shown in this paper since they are anything but interesting. The strategy of our parametrization is laid out as follows.

(I) We propose two {\it basic} fragmentation functions $D_{B}^{\rm val} (z)$ and $D_{B}^{\rm sea} (z)$ which represent the valence and sea contributions to the prompt baryon fragmentation function. They can be obtained by solving the following equations
\begin{align}
&
D_{1,u}^{\Lambda,{\rm dir}} (z) = \lambda_s [D_{B}^{\rm val} (z) + D_{B}^{\rm sea} (z)] = \frac{1+z}{2} D_{1,u}^{\Lambda/\bar \Lambda} (z), 
\\
&
D_{1, \bar u}^{\Lambda,{\rm dir}} (z) = \lambda_s D_{B}^{\rm sea} (z) = \frac{1-z}{2} D_{1, u}^{\Lambda/\bar \Lambda} (z), 
\end{align}
where $\lambda_s=1/3$ is the strangeness suppression factor and $D_{1,u}^{\Lambda/\bar \Lambda} (z)$ is given by the DSV parametrization~\cite{deFlorian:1997zj} with the factorization being specified by $\mu_f=1$ GeV. Throughout this paper, we neglect the QCD evolution effect. 

(II) We also introduce $\kappa = 1/3$ as the production ratio of $J^P = (3/2)^+$ and $J^P = (1/2)^+$ baryons. It is thus straightforward to obtain prompt fragmentation functions of all different baryons. For instance, we have
\begin{align}
& 
D_{1,u/d}^{\Lambda,{\rm dir}} = \lambda_s [D_{B}^{\rm val} (z) + D_{B}^{\rm sea} (z)],
\\
&
D_{1,s}^{\Lambda,{\rm dir}}   = D_{B}^{\rm val} (z) + \lambda_s D_{B}^{\rm sea} (z),
\\
& 
D_{1,u}^{\Xi^-,{\rm dir}} (z) = \lambda_s^2 D_B^{\rm sea} (z),
\\
& 
D_{1,d}^{\Xi^-,{\rm dir}} (z) = \lambda_s^2 [D_B^{\rm val} (z) + D_B^{\rm sea} (z)],
\\
&
D_{1,s}^{\Xi^-,{\rm dir}} (z) = 2\lambda_s D_B^{\rm val} (z) + \lambda_s^2 D_B^{\rm sea} (z),
\\
& 
D_{1,u/d}^{\Omega^-, {\rm dir}} (z) =  \lambda_s^3 \kappa D_B^{\rm sea} (z),
\\
&
D_{1,s}^{\Omega^-, {\rm dir}} (z) =  3\lambda_s^2 \kappa D_B^{\rm val} (z) + \lambda_s^3 \kappa D_B^{\rm sea} (z).
\end{align}
Parametrizations of other baryons follow the same rules and are not explicitly shown in this paper.

\begin{figure}[htb]
\centering
\includegraphics[width=0.45\textwidth]{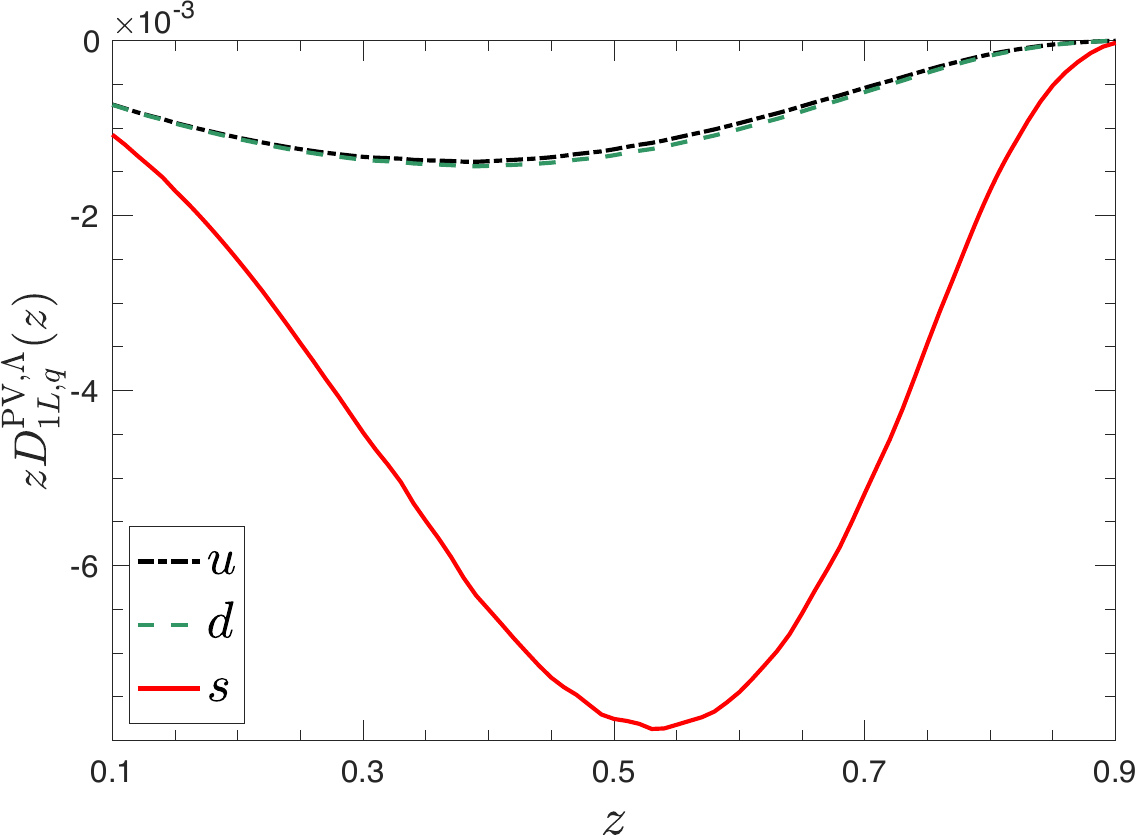}
\includegraphics[width=0.45\textwidth]{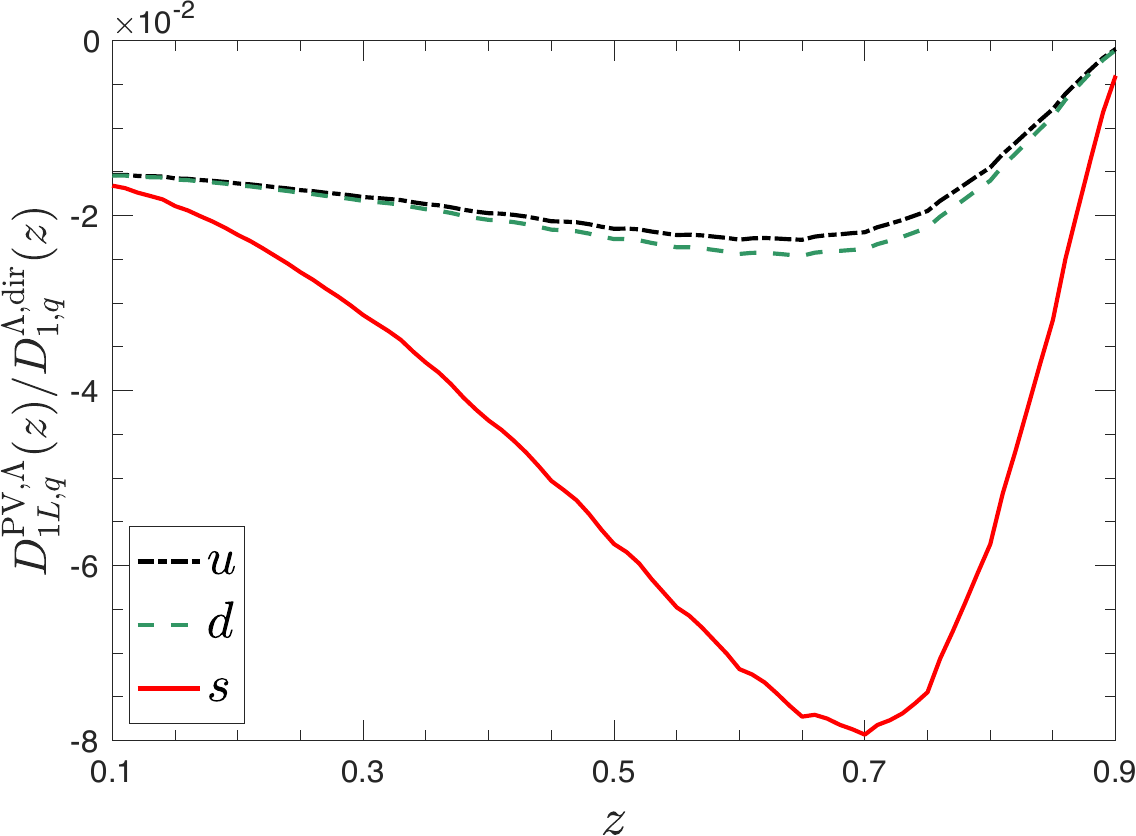}
\caption{Numerical estimate of the P-odd fragmentation function $D_{1L,q}^{{\rm PV},\Lambda}(z)$.}
\label{fig:Dtilde_Lambda}
\end{figure}

Convoluting with the kernel function, we obtain the decay contribution to the P-odd FFs. The numerical results for $D_{1L,q}^{{\rm PV},\Lambda} (z)$ are shown in the l.h.s. of Fig.~\ref{fig:Dtilde_Lambda}. We see that the fragmentation function for $s$-quark, $D_{1L,s}^{{\rm PV},\Lambda} (z)$, is much larger than those for $u$- and $d$-quarks. This is because $D_{1L,q}^{{\rm PV},\Lambda} (z)$'s are completely generated from hadron weak decay. For the decay contribution of $\Lambda$ production, the weak decay from $\Xi$ baryons is dominant. Also notice that the FFs of $u$- or $d$-quark to $\Xi$ are much smaller than that for $s$-quark due to the strangeness suppression and the valance contribution, it is not surprising that $D_{1L,u/d}^{{\rm PV},\Lambda} (z)$ are much smaller than $D_{1L,s}^{{\rm PV},\Lambda} (z)$. To illustrate the relative magnitude of the P-odd FFs, we also draw the ratios of $D_{1L,q}^{{\rm PV},\Lambda} (z)$ to the $\Lambda$ direct FFs in the r.h.s. of Fig.~\ref{fig:Dtilde_Lambda}. We see especially for $s$-quark that the ratio can reach $\sim 8\%$ which is not tiny. This indicates that the P-odd FFs may have observable effects.

\section{Observables}

In this section, we present our numerical estimates concerning two observables resulting from P-odd fragmentation functions. 

\subsection{Longitudinal polarization of $\Lambda$ hyperons produced in $e^+ e^-$ annihilations}

As mentioned above, $D_{1L}^{\rm PV} (z)$ is responsible for generating longitudinally polarized hadrons from unpolarized partons. Therefore, we expect that there will be a signal of spontaneous longitudinal polarization for the $\Lambda$-hyperons produced in unpolarized high-energy collisions. The produced $\Lambda$-hyperons can further decay into the $p+\pi^-$ pair through the weak interaction. The polarization of the proton is usually averaged over since it is extremely difficult to measure with current experimental instruments. Eventually, the angular distribution of the final state proton follows the $dN/d\Omega \propto 1 + \alpha {\cal P}_L \cos \theta^*$ in the $\Lambda$-rest frame. Here, ${\cal P}_L$ is the polarization of $\Lambda$, $\alpha$ is a constant, and $\theta^*$ is the polar angle of the proton momentum in the $\Lambda$-rest frame. Therefore, the $\Lambda$ polarization, ${\cal P}_L$ can be easily measured by extracting $\langle \cos\theta^* \rangle$ through its self-analyzing decay.

Quarks produced in $e^+e^-$ annihilation can acquire longitudinal polarization through the $Z^0$-boson exchange diagram and the interference term. The quark longitudinal polarization is significant at the $Z^0$-pole and it is negligible at relatively low energy. Eventually, this longitudinal polarization of quark can be inherited by the fragmented hadron through $G_{1L}$ which is a P-even fragmentation function and becomes the background of our study. The second part of contribution comes from P-odd fragmentation functions. At this point, we take a brief detour to the $pp$ collisions. In $pp$ collisions, the partonic collisions are dominated by the strong interaction. The fragmenting hadron is not polarized. Therefore, the unpolarized $pp$ collisions provide a better platform to probe the P-odd fragmentation functions. However, in this work, we still utilize the $e^+e^-$ annihilation process as an example to demonstrate this effect. Particularly, the ratio between these two terms in Eq.~(\ref{eq:polarization}) offers information on the magnitude of P-odd effects. The final contribution to the longitudinal polarization of $\Lambda$ in $e^+e^-$ annihilation reads
\begin{align}
{\cal P}_L (z) = \frac{\sum_q \left[ \Delta \omega_q G_{1L,q} (z) + \omega_q D_{1L,q}^{\rm PV} (z) \right]}{\sum_q \omega_q D_{1,q} (z)}. \label{eq:polarization}
\end{align}
Here, $\omega_q = e_q^2 + c_1^e c_1^q \chi + \chi_{\rm int}^q c_V^e c_V^q$ is the weight function for the unpolarized quark production and $\Delta \omega_q = - c_1^e c_3^q \chi - c_V^e c_A^q \chi_{\rm int}^q$ with $c_1^e = (c_V^e)^2 + (c_A^e)^2$, $c_1^q = (c_V^q)^2 + (c_A^q)^2$, $c_3^q = 2c_V^q c_A^q$, $\chi=Q^4 /[(Q^2-M_Z^2)^2+\Gamma_Z^2 M_Z^2] \sin ^4 2 \theta_W$, $\chi_{\text {int }}^q=-2 e_q Q^2(Q^2-M_Z^2) /[(Q^2-M_Z^2)^2+\Gamma_Z^2 M_Z^2] \sin ^2 2 \theta_W$. $M_Z$ is the mass of $Z^0$-boson, $\Gamma_Z$ is the width of the $Z^0$-boson mass, $\theta_W$ is the Weinberg angle and $Q$ is the center-of-mass energy of the colliding leptons. For simplicity, we have only shown the contribution from quark to $\Lambda$. The contribution from antiquark is small but should be taken into account. However, notice that $\omega_{\bar q} = \omega_{q}$ and $\Delta \omega_{\bar q} = - \Delta \omega_q$. The first one comes from the simple fact that the cross section of quark production is the same with that of antiquark production. The second one indicates that quark and antiquark have the opposite helicities.

\begin{figure}[htb]
\includegraphics[width=0.45\textwidth]{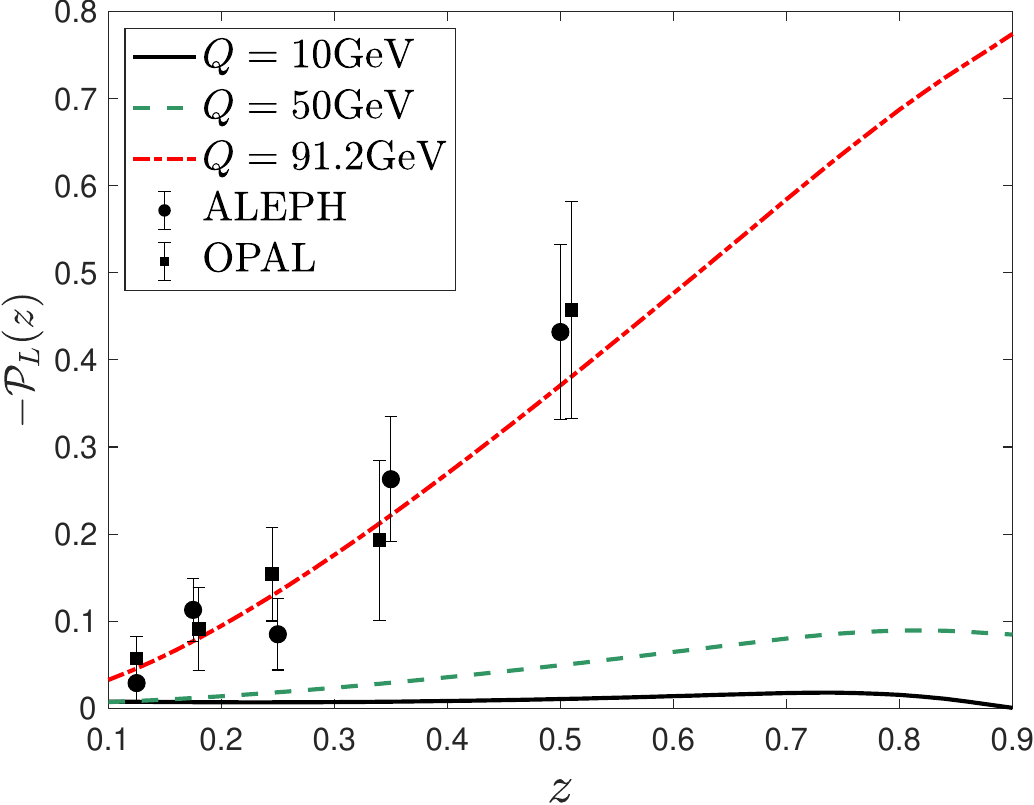}
\includegraphics[width=0.45\textwidth]{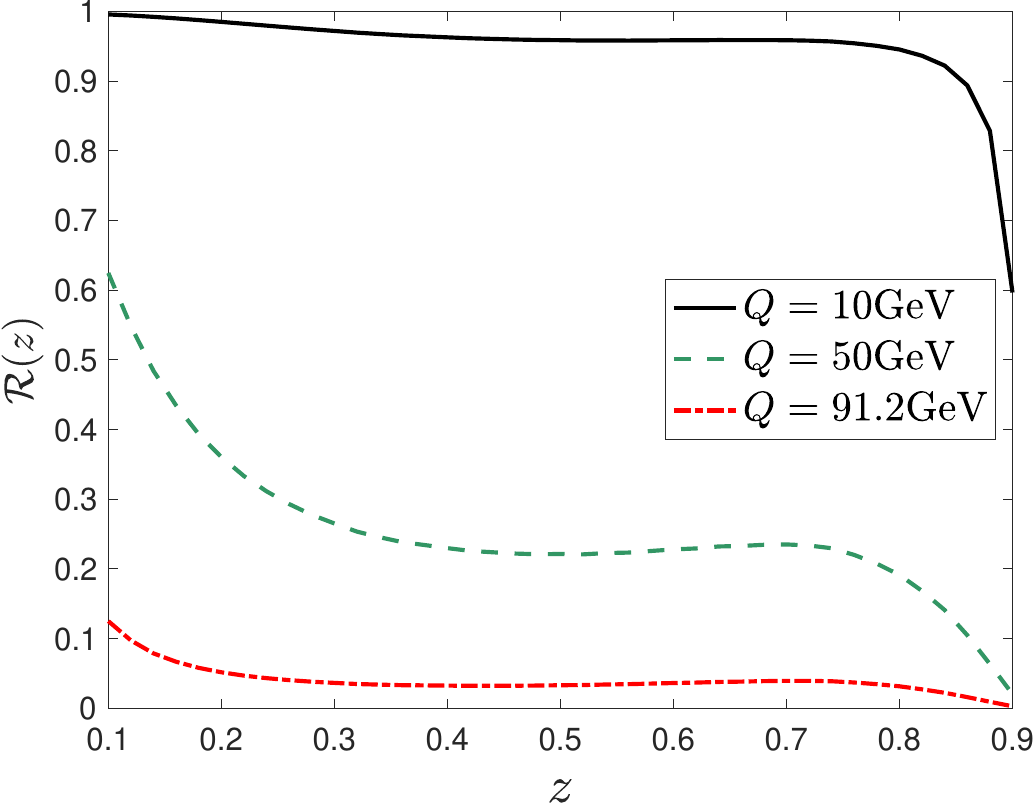}
\caption{LHS: Polarization of $\Lambda$ hyperons produced in $e^+e^-$ annihilation at different energies. RHS: Relative contribution of the P-odd fragmentation function at different collisional energies.
\label{fig:lambda-pol}}
\end{figure}

To proceed, a parameterization of the longitudinal spin transfer $G_{1L,q} (z)$ is required. We take $G_{1L,q}^{\Lambda}(z) = z^{0.8} D_{1,q}^{\Lambda}(z)$ and only consider contributions from light quarks. Such a simple parameterization can already provide a sufficiently good description of the $\Lambda$-hyperon polarization measured by the LEP experiment~\cite{ALEPH:1996oew, OPAL:1997oem}. With these approximations at hand, we present our numerical result for the longitudinal polarization of $\Lambda$ hyperons at different collision energies in l.h.s. of Fig.~\ref{fig:lambda-pol}. To illustrate the relative contribution from the P-odd fragmentation function, we define the ratio ${\cal R} (z)$ as
\begin{align}
{\cal R} (z) = \frac{|\sum_q \omega_q D_{1L,q}^{\rm PV} (z)| }{|\sum_q \Delta \omega_q G_{1L,q} (z)| + |\sum_q \omega_q D_{1L,q}^{\rm PV} (z)|},
\end{align}
and show the corresponding numerical results in r.h.s. of Fig.~\ref{fig:lambda-pol}. It is obvious that the dominant contribution to the hadron polarization arises from the P-odd effect when $Q \ll M_Z$. The relative contribution from $D_{1L}^{\rm PV}$ decreases as the collisional energy increases and reaches the minimum at the $Z^0$-pole. Thus, it is better to access this P-odd fragmentation function at relatively low energy collisions. 

\begin{figure}
\includegraphics[width=0.45\textwidth]{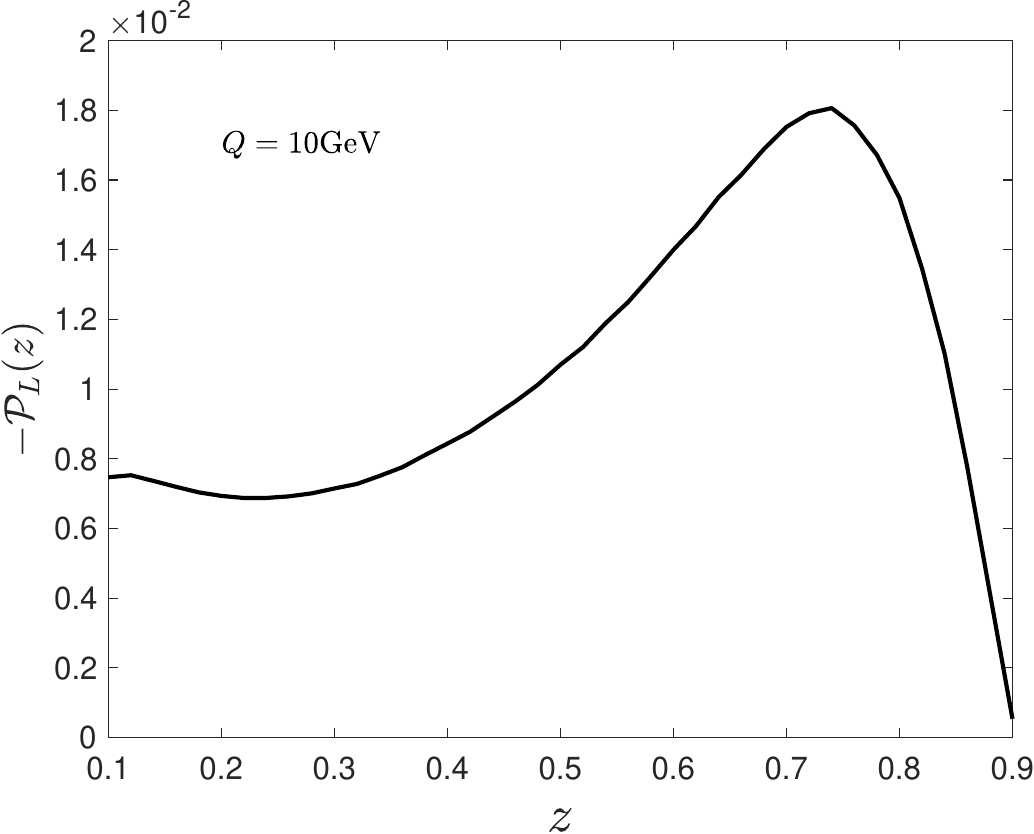}
\includegraphics[width=0.45\textwidth]{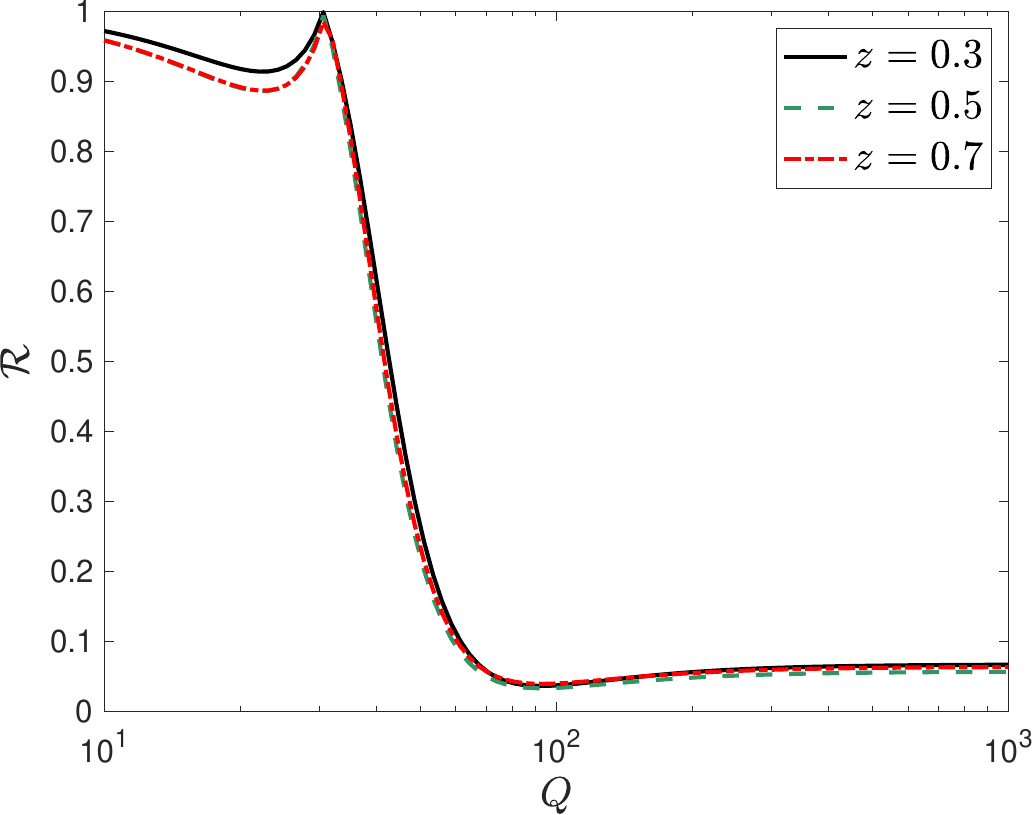}
\caption{LHS: Polarization of $\Lambda$ hyperons produced in $e^+e^-$ annihilation at $\sqrt{s}=10$ GeV. RHS: relative contribution from the P-odd fragmentation function as a function of collisional energy at different $z$.
\label{fig:q-dependence}}
\end{figure}

Furthermore, we show the longitudinal polarization of the $\Lambda$ hyperons produced in $e^+e^-$ annihilations at $\sqrt{s}=10$ GeV in the l.h.s. of Fig.~\ref{fig:q-dependence} and the relative contribution from the P-odd fragmentation function as a function of collisional energy at different $z$ in the r.h.s. of Fig.~\ref{fig:q-dependence}. Notice that the ratio ${\cal R}$ even reaches unity at about $Q\sim 30$ GeV, which indicates that the contribution from the spin transfer of polarized quarks vanishes. The reason is straightforward. Both the $Z^0$-boson exchange term and the $\gamma Z^0$ interference term contribute to the quark polarization. However, the signs are different for $d$ and $s$ quarks. Namely, the interference term contributes to the positive polarization while the $Z^0$-boson exchange diagram contributes to the negative. At low-$Q^2$, the dominant contribution comes from the interference term, which results in positive polarization for the final state $\Lambda$ hyperon. However, at higher energy, $Z^0$-boson exchange diagram takes charge and eventually turns the polarization of $\Lambda$-hyperons into negative. Therefore, at a certain energy scale, there is a cross-over where the P-even contribution vanishes.

\subsection{Modification to the back-to-back dihadron production rate}

Although quarks produced through electromagnetic interaction in the $e^+e^-$-annihilation process are not polarized individually, the polarization correlation of quark and antiquark takes the maximum value \cite{Chen:1994ar, Zhang:2023ugf}. The polarization correlation is thus proposed as a new probe for the $G_{1L}$ fragmentation function. In this work, we focus on the P-odd effect of this polarization correlation of the quark-antiquark pair. 

Combining the P-even and P-odd contributions, the complete cross section of spin-0 dihadron production in $e^+e^-$ annihilation process is given by~\cite{Boer:1997mf}
\begin{align}
\frac{d \sigma(e^+e^-\to h_1 h_2 X)}{dz_{h_1} dz_{h_2}} = \frac{4\pi \alpha_{\rm em}^2}{Q^2} \sum_{q} 
&
\omega_q \bigl[ D_{1,q}^{h_1}(z_{h_1}) D_{1,\bar q}^{h_2} (z_{h_2}) - G_{1,q}^{{\rm PV},h_1}(z_{h_1}) G_{1,\bar q}^{{\rm PV},h_2} (z_{h_2})  \bigr]
\nonumber\\
+
&
\Delta \omega_q \bigl[G_{1,q}^{{\rm PV},h_1}(z_{h_1}) D_{1,\bar q}^{h_2} (z_{h_2}) - D_{1,q}^{h_1}(z_{h_1}) G_{1,\bar q}^{{\rm PV},h_2} (z_{h_2}) )  \bigr]
, \label{eq:cs-ee}
\end{align}
where $\omega_q$ and $\Delta\omega_q$ have already defined in the previous subsection. Again, although we only lay out the contribution from $q\to h_1$ and $\bar q \to h_2$ here, the exchange between quark and antiquark is implicit.

The first term in Eq.~(\ref{eq:cs-ee}) computes P-even contributions with the other terms evaluating P-odd ones. The emergence of P-odd terms clearly modifies the production rate of dihadron production. Depending on the hadron species, the modification could be a suppression or an enhancement. The magnitude of this modification can be quantified by ${\cal M} (z_{h_1},z_{h_2})$ which is defined as
\begin{align}
{\cal M} (z_{h_1}, z_{h_2}) \equiv \frac{\sum_q \omega_q G_{1,q}^{{\rm PV},h_1}(z_{h_1}) G_{1,\bar q}^{{\rm PV},h_2} (z_{h_2}) - \sum_q \Delta\omega_q \bigl[G_{1,q}^{{\rm PV},h_1}(z_{h_1}) D_{1,\bar q}^{h_2} (z_{h_2}) - D_{1,q}^{h_1}(z_{h_1}) G_{1,\bar q}^{{\rm PV},h_2} (z_{h_2}) )  \bigr]}{\sum_q \omega_q D_{1,q}^{h_1}(z_{h_1}) D_{1,\bar q}^{h_2} (z_{h_2})}.
\label{eq:dihadron}
\end{align}
The sign of $\cal M$ indicates if the modification is a suppression or an enhancement. If $\cal M$ is positive, we have a suppression. Otherwise, we have an enhancement.

\begin{figure}[htb]
\includegraphics[width=0.3\textwidth]{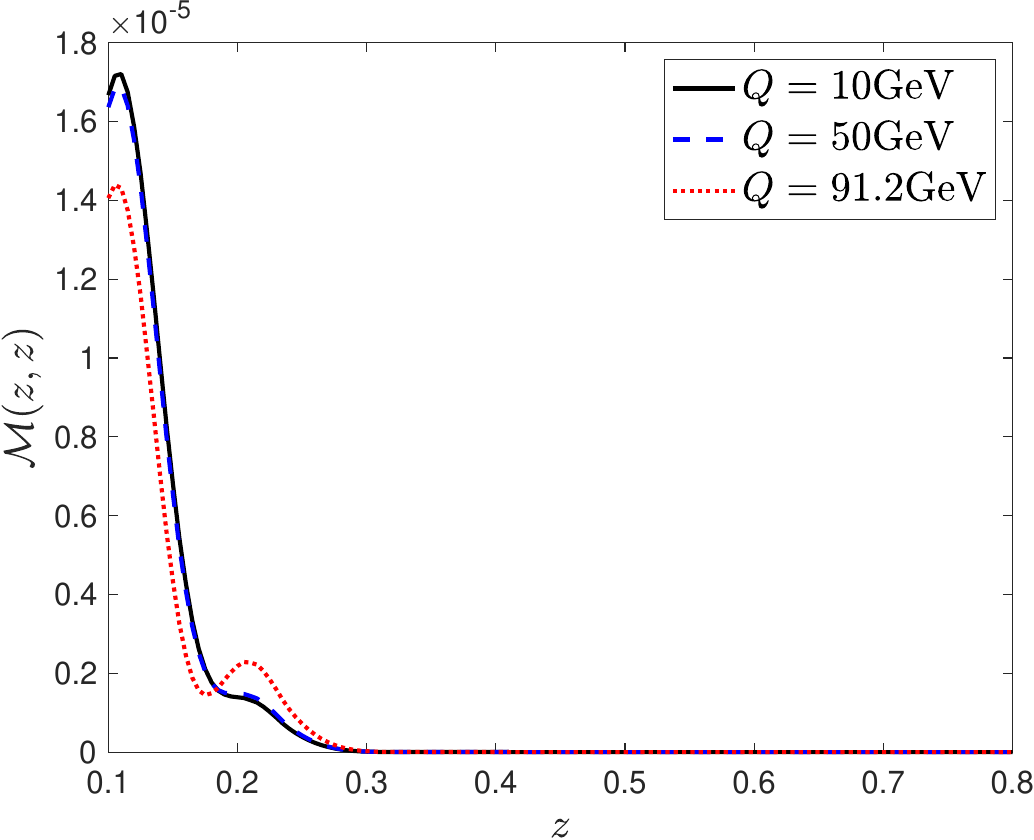}
\includegraphics[width=0.3\textwidth]{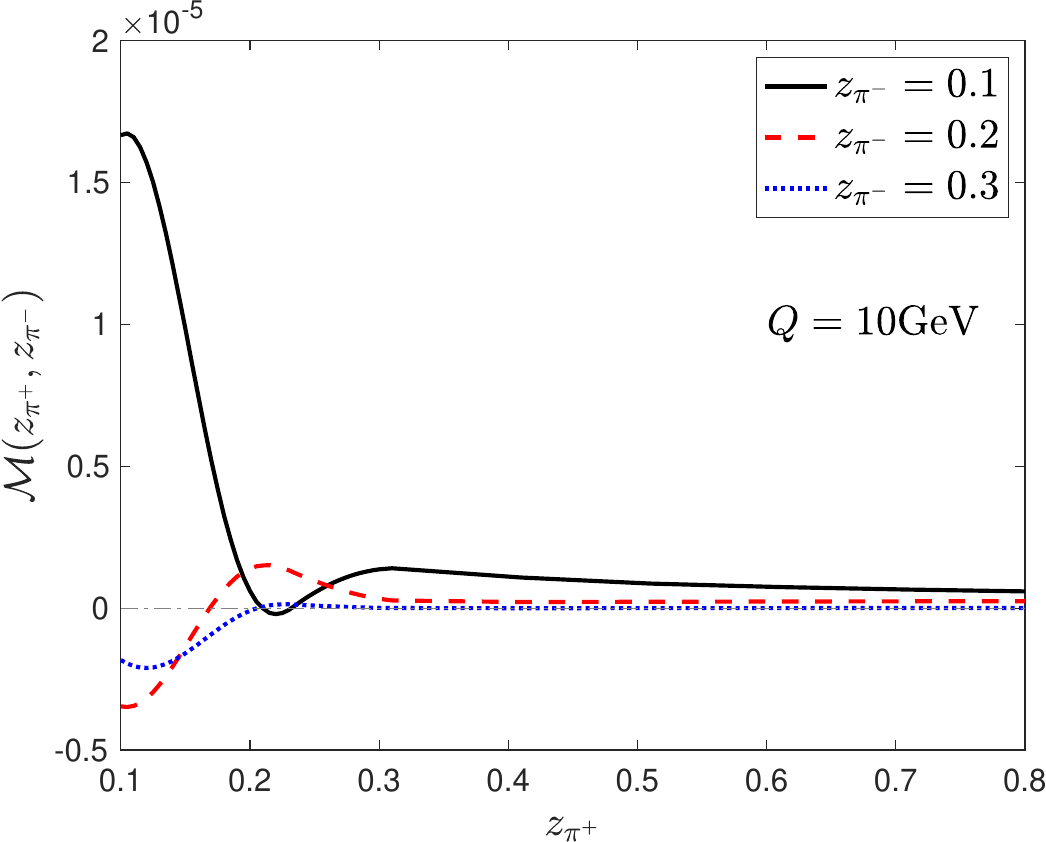}
\includegraphics[width=0.3\textwidth]{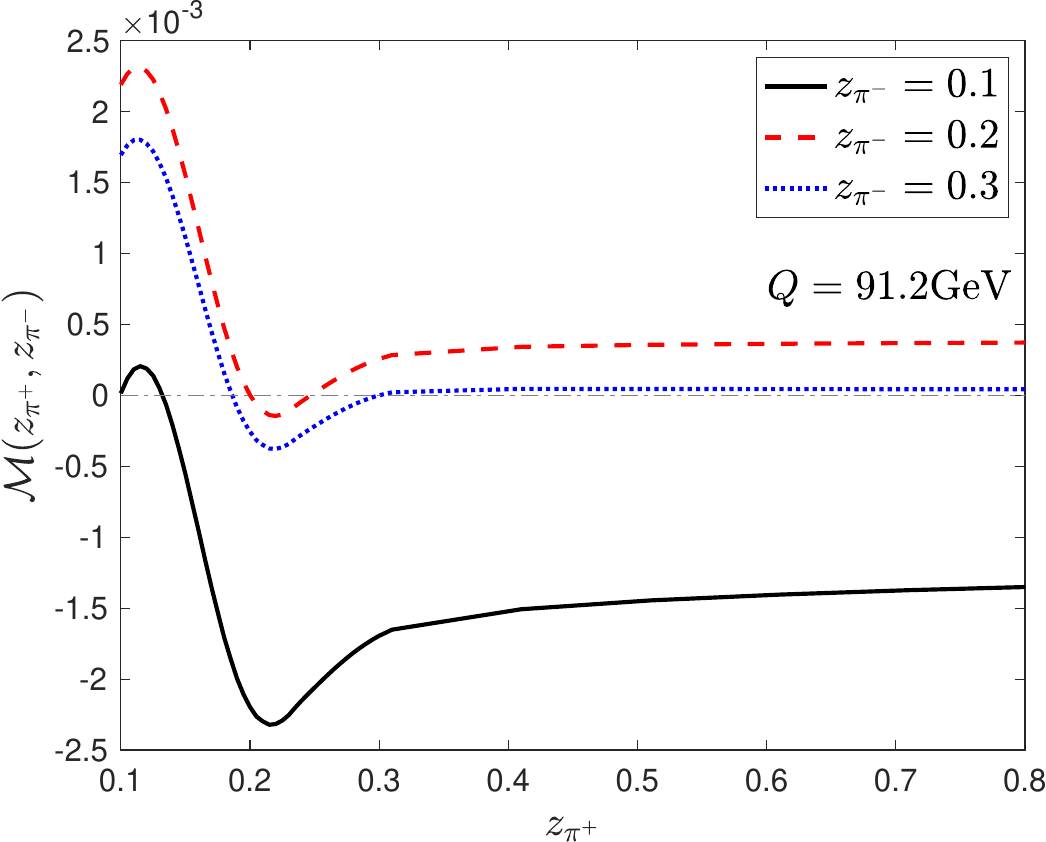}
\caption{Left: Suppression of $\pi^+\pi^-$ production in $e^+e^-$ annihilation as a function of $z = z_{\pi^+} = z_{\pi^-}$; Middle: Modification of $\pi^+\pi^-$ production in $e^+e^-$ annihilation as a function of $z_{\pi^+}$ at $\sqrt{s}=10$ GeV; Right: Modification of $\pi^+\pi^-$ production in $e^+e^-$ annihilation as a function of $z_{\pi^+}$ at $\sqrt{s}=91.2$ GeV.}
\label{fig:dihadron}
\end{figure}

We consider the $\pi^+ \pi^-$-pair production and first show our numerical estimate of ${\cal M} (z_{\pi^+}, z_{\pi^-})$ at $z_{\pi^+} = z_{\pi^-} =z$ as a function of $z$ in the left panel of Fig.~\ref{fig:dihadron}. Effectively, the last two terms in Eq. (\ref{eq:dihadron}) largely cancel each other due to the charge conjugation symmetry by choosing $z_{\pi^+}=z_{\pi^-}$. Therefore, we find ${\cal M} (z,z) \simeq \sum_q \omega_q [G_{1,q}^{\rm PV} (z)]^2 / \sum_q\omega_q[D_{1,q} (z)]^2$ which is positive definite. The ratio $|G_{1,q}^{\rm PV} (z)/D_{1,q} (z)|$ is at the order of $10^{-2}$, which explains why the signal is tiny. As long as $z_{\pi^+}$ and $z_{\pi^-}$ are not the same, the polarized quark term contributes. This brings a significant parton energy dependence since the quark polarization is negligible at low $Q$ and takes the maximum at $Q=M_Z$. The negligible parton polarization at the low $Q$ region results in a tiny modification of dihadron pair production. However, the magnitude of the modification becomes 100 times larger at around the $Z^0$-pole. This feature is shown in the middle and right panels of Fig.~\ref{fig:dihadron}. 

In light of the above discussion, we recommend the following kinematic region to minimize the weak decay contribution and thus to better probe the strong-parity-violating effect. (1) Measure the $\pi^{+}\pi^{-}$-pair production in low-$Q$ $e^+e^-$ experiments; (2) Focus on the large $z$ region, which strongly suppresses the weak decay contribution; (3) Set $z_{\pi^+}$ and $z_{\pi^-}$ to be equal to minimize the contribution from polarized quarks.

\section{Summary}

Two sources can contribute to the P-odd fragmentation functions. The first one is the QCD $\theta$ term. The second one is the weak decay of heavier hadrons. While the first source can be employed to study the tiny $\theta$ parameter, the second one becomes an important background in such a study. A good understanding of the decay contribution is required to constrain the tiny $\theta$ parameter. 

We make a systemic investigation of the decay contribution to the P-odd fragmentation functions and establish the general formula to compute the decay contribution of fragmentation functions. P-odd fragmentation functions lead to several phenomena that are prohibited by parity conservation such as the spontaneous longitudinal polarization of spin-$1/2$ hadrons produced from unpolarized quarks and the modification of the dihadron production rate in $e^+e^-$ annihilation process. 

We employ a simple parametrization to estimate the magnitude of these P-odd fragmentation functions for $\Lambda$ and pions and find that typically they are about a few percent of the corresponding unpolarized fragmentation function. As a result, $\Lambda$-hyperons produced from unpolarized partons gain a longitudinal polarization as large as one or two percent. Such an effect can also be observed in other high-energy collisions such as unpolarized pp or AA collisions. In light of this, the impact should be rigorously reviewed particularly for the polarization observables with tiny magnitudes. 

We also investigate the modification of dihadron pair production. The numerical results show that the modification is small in low-$Q$ collisions. However, it increases to the order of $10^{-3}$ at around the $Z^0$-pole. Based on this calculation, we recommend that the large $z_{\pi^+} \simeq z_{\pi^-}$ region in low-$Q$ $e^+e^-$ collisions is the optimal place to study the strong parity violating effect. 

\begin{acknowledgments}

We thank Z.T. Liang for inspiring discussions. This work is supported in part by the National Natural Science Foundation of China (approval number 12375075, 11505080, 12005122), the Taishan fellowship of Shandong Province for junior scientists, the Shandong Province Natural Science Foundation under grant No. 2023HWYQ-011, No. ZR2018JL006, and No. ZR2020QA082, and Youth Innovation Team Program of Higher Education Institutions in Shandong Province (Grant No. 2023KJ126).

\end{acknowledgments}

\end{document}